\documentclass[aps,prl,twocolumn,groupedaddress,showpacs]{revtex4}

\usepackage{graphicx,amsmath}
\usepackage{feynmp}
\usepackage{mathrsfs}
\usepackage{gensymb}
\usepackage{subfigure}
\usepackage{amsfonts}
\usepackage{times}
\usepackage{setspace}
\usepackage{color}

\begin{document}

\title{Identifying  topological order in the Shastry-Sutherland model via entanglement entropy}

\author{David C. Ronquillo and Michael R. Peterson}
\affiliation{Department of Physics and Astronomy, California State University Long Beach,  Long Beach, California 90840, USA}

\begin{abstract}
\begin{spacing}{1.0593}
It is known that for a topologically ordered state the area law for the entanglement entropy shows a negative universal additive constant contribution, $-\gamma$, called the topological entanglement entropy.  We theoretically study the entanglement entropy of the two-dimensional Shastry-Sutherland quantum antiferromagnet using exact diagonalization on clusters of 
16 and 24 spins.  By utilizing the Kitaev-Preskill construction [A. Kitaev and J. Preskill, Phys. Rev. Lett. {\bf 96}, 110404 (2006)] we extract a finite topological term, $-\gamma$, in the region of bond-strength parameter space corresponding to high geometrical frustration.  Thus, we provide strong 
evidence for the  existence of an exotic topologically ordered state  and shed light on the nature of this model's strongly frustrated, and long controversial, intermediate phase.
\end{spacing}
\end{abstract}

\date{\today}

\pacs{75.10.Jm, 03.65.Ud, 75.10.Kt}

\maketitle

\textit{Introduction}. Much attention has been given to the ground-state properties of two-dimensional (2D) quantum antiferromagnets in recent decades. Of particular interest to both experimentalists and theorists are the possible exotic phases of matter that may arise at zero temperature within given regions of these system's bond-strength parameter space. By tweaking the relative strengths of their bonds, systems may be induced into highly frustrated states whose physics becomes determined by quantum fluctuations.

One such 2D quantum antiferromagnet whose parameter space has received considerable attention is well described by the Shastry-Sutherland model (SSM) \cite{Shastry1981, Albrecht1996, Miyahara1999, Weihong1999, Koga2000, Carpentier2001, Chung2001, Zheng2001, Lauchli2002, Miyahara2003, Hajj2005, Darradi2005, Isacsson2006, Moukouri2008, Kim2013, Corboz2013}, whose Hamiltonian is given by 
\begin{eqnarray}
H = J_{1}\sum_{\langle i,j\rangle}{\bf{S}}_{i}\cdot{\bf{S}}_{j}+J_{2}\sum_{\langle l,m\rangle}{\bf{S}}_{l}\cdot{\bf{S}}_{m}.
\label{Ham}
\end{eqnarray}
This deserved scrutiny is due to the SSM's inherent theoretical interest as well as its connection to the fascinating experimental system SrCu$_2$(BO$_3$)$_2$, whose magnetization curve displays quantized plateaus \cite{Miyahara1999, Kageyama99,Misguich2001,Carpentier2001,Chung2001,Sebastian2008}.  For $J_1,J_2>0$, Eq. (\ref{Ham}) describes antifferomagnetic interactions between $N$ spin-1/2 degrees of freedom lying on a square lattice having $J_{1}$ strength axial bonds, and $J_{2}$ strength diagonal bonds on every other square (see Fig.~\ref{16kp}). Here $\langle i,j\rangle$ and $\langle l,m\rangle$ refer to axially and diagonally connected degrees of freedom, respectively.

In their original paper, Shastry and Sutherland (SS) rigorously showed that the ground-state wave function of Eq.(\ref{Ham}) is a product of singlets along all diagonal bonds 
\begin{equation}\label{dimerprod}
|\psi_{g}\rangle=\prod_{\langle l,m\rangle}\frac{1}{\sqrt{2}}(|\uparrow\downarrow\rangle-|\downarrow\uparrow\rangle)_{l,m},
\end{equation}
for $g\equiv J_{2}/J_{1}\geq2$, with energy per site of $E_{0}/N=-(3/8)g$ \cite{Shastry1981, Albrecht1996}.  In fact, Eq. (\ref{dimerprod}) continues to represent an eigenstate of $H$ across all parameter space \cite{Shastry1981}. This dimer-singlet valence bond state is gapped to its first triplet excitation and has no long-range magnetic order \cite{Shastry1981, Miyahara1999, Weihong1999}. Further studies of the SSM's parameter space have revealed that Eq. (\ref{dimerprod}) remains the  ground-state wave function for values of $g$ lying well below the rigorously demonstrated $g=2$ \cite{Albrecht1996, Miyahara1999, Weihong1999, Koga2000, Carpentier2001, Chung2001, Zheng2001, Lauchli2002, Miyahara2003, Hajj2005, Darradi2005, Isacsson2006, Moukouri2008, Kim2013, Corboz2013}.  

For $g\ll1$, the SSM is N\'{e}el ordered due to relatively weak diagonal interactions \cite{Miyahara2003, Moukouri2008}. Between this magnetic long-range order and the dimer-singlet valence bond solid (VBS) order, described by Eq.~(\ref{dimerprod}), lies a region of high frustration whose exact nature has eluded certain classification for almost 20 years. 

\begin{figure}
\begin{center}
\hspace*{-.05cm}
\includegraphics[width=8.5cm, height=6.4cm, angle=0]{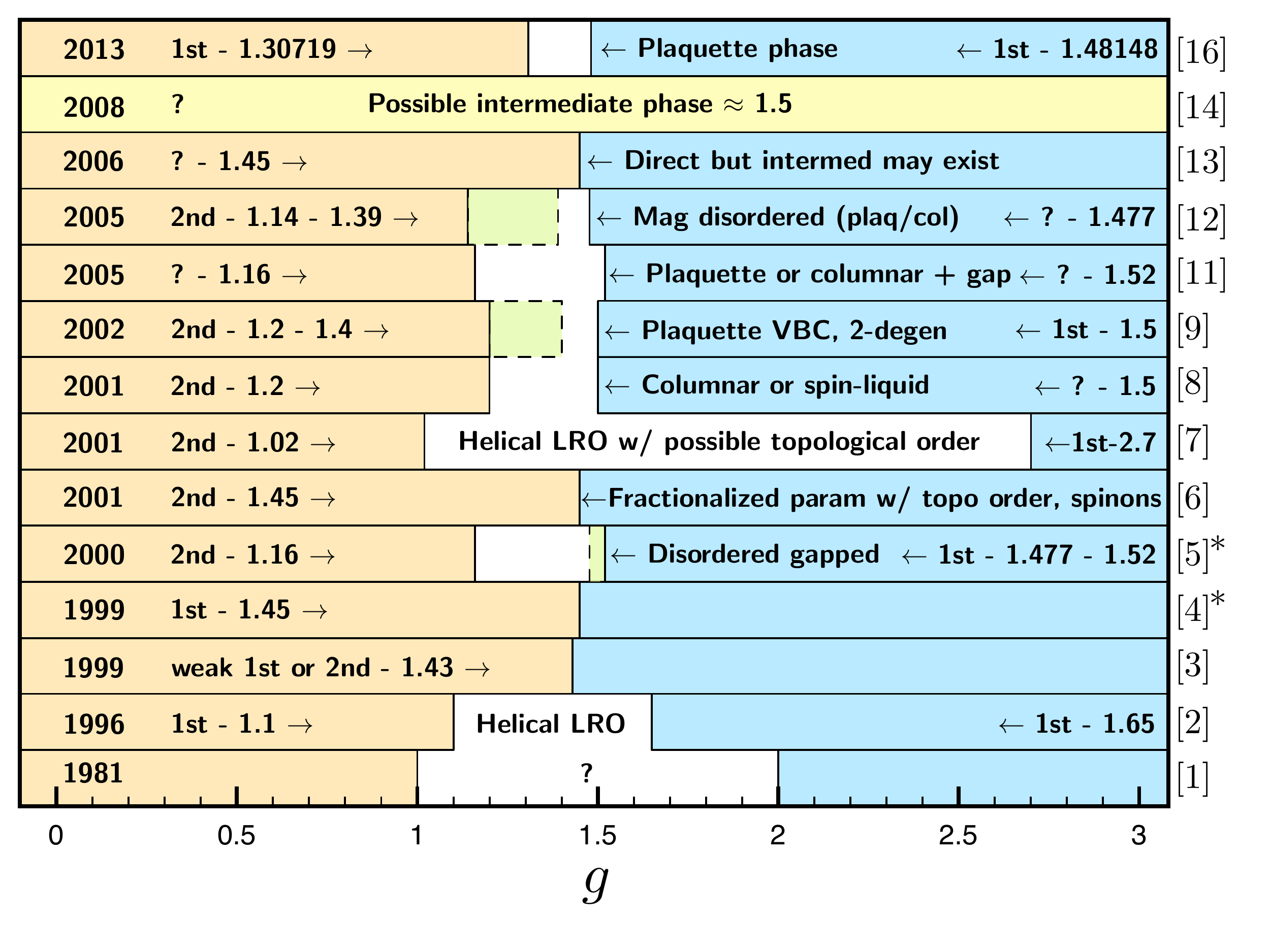}
\caption{(Color online) A summary of some important prior contributions evaluating the quantum phase diagram of the Shastry-Sutherland model. From left to right, each horizontal bar provides the following information: (i) year research was conducted, (ii) order of phase transition between N\'{e}el and intermediate phase, (iii) proposed location in parameter space of latter transition point, (iv) type of order proposed for intermediate phase, (v) order of phase transition between intermediate phase and the Shastry-Sutherland dimer-singlet phase, and (vi) proposed location in parameter space of latter transition point. References are provided towards the rightmost edge of this graphic. * studies incorporate analysis of the experimental system SrCu$_{2}$(BO$_{3}$)$_{2}$ whose behavior and properties are well described by the Shastry-Sutherland model.}
\label{Fig1}
\end{center}
\end{figure}

Figure \ref{Fig1} summarizes some important prior contributions to the classification effort. Various numerical methods have been employed, including linear spin-wave theory \cite{Albrecht1996}, Schwinger boson mean-field theory \cite{Albrecht1996}, series expansion method \cite{Weihong1999, Koga2000, Zheng2001}, effective mean-field theory \cite{Carpentier2001}, bond operator method \cite{Lauchli2002}, perturbation and renormalized excitonic method \cite{Hajj2005}, high-order coupled cluster method \cite{Darradi2005}, variational algorithm based on projected entangled pair states (PEPS) \cite{Isacsson2006}, two-step density-matrix renormalization group (DMRG) \cite{Moukouri2008}, variational algorithm based on infinite projected entangled-pair states (iPEPS) \cite {Corboz2013}, and exact diagonalization on various sized clusters \cite{Albrecht1996, Miyahara1999, Koga2000, Lauchli2002}. Most authors have discerned the existence of an intermediate phase lying somewhere between $1.14<g<1.5$. While its exact nature has been controversial, it is likely to be gapped \cite{Koga2000, Lauchli2002, Hajj2005} and magnetically disordered \cite{Albrecht1996, Koga2000, Lauchli2002, Hajj2005, Darradi2005, Moukouri2008}.

Greater insight into the nature of the SSM's intermediate phase has been stifled by, what some have acknowledged as, the inadequacy of the available tools used for probing the relevant region of the  SSM's parameter space \cite{Zheng2001, Miyahara2003, Isacsson2006}. Here we shed light on the nature of the intermediate phase by resorting to a new tool, the topological entanglement entropy, whose service in this effort has, until this work, been neglected. With topological entanglement entropy we show that the SSM is topologically ordered within a finite region of its parameter space. The consequences of this result may prove relevant to those interested in the possibility of developing a physically fault-tolerant quantum computing device \cite{Nayak2008}. 

\textit{Review and motivations}. In their field theoretic work Carpentier and Balents \cite{Carpentier2001}, and Chung, Marston, and Sachdev \cite{Chung2001} have suggested that for a narrow range in $g$ (near the point of maximal frustration) the ground state of the SSM exhibits a degeneracy which is topological in origin. An appeal to numerical methods for exploring this frustrated regime is explicitly made, which we currently supply.

\begin{figure}
\vspace*{.03cm}
\begin{center}
\hspace*{-.65cm}
\includegraphics[width=8.5cm, angle=0]{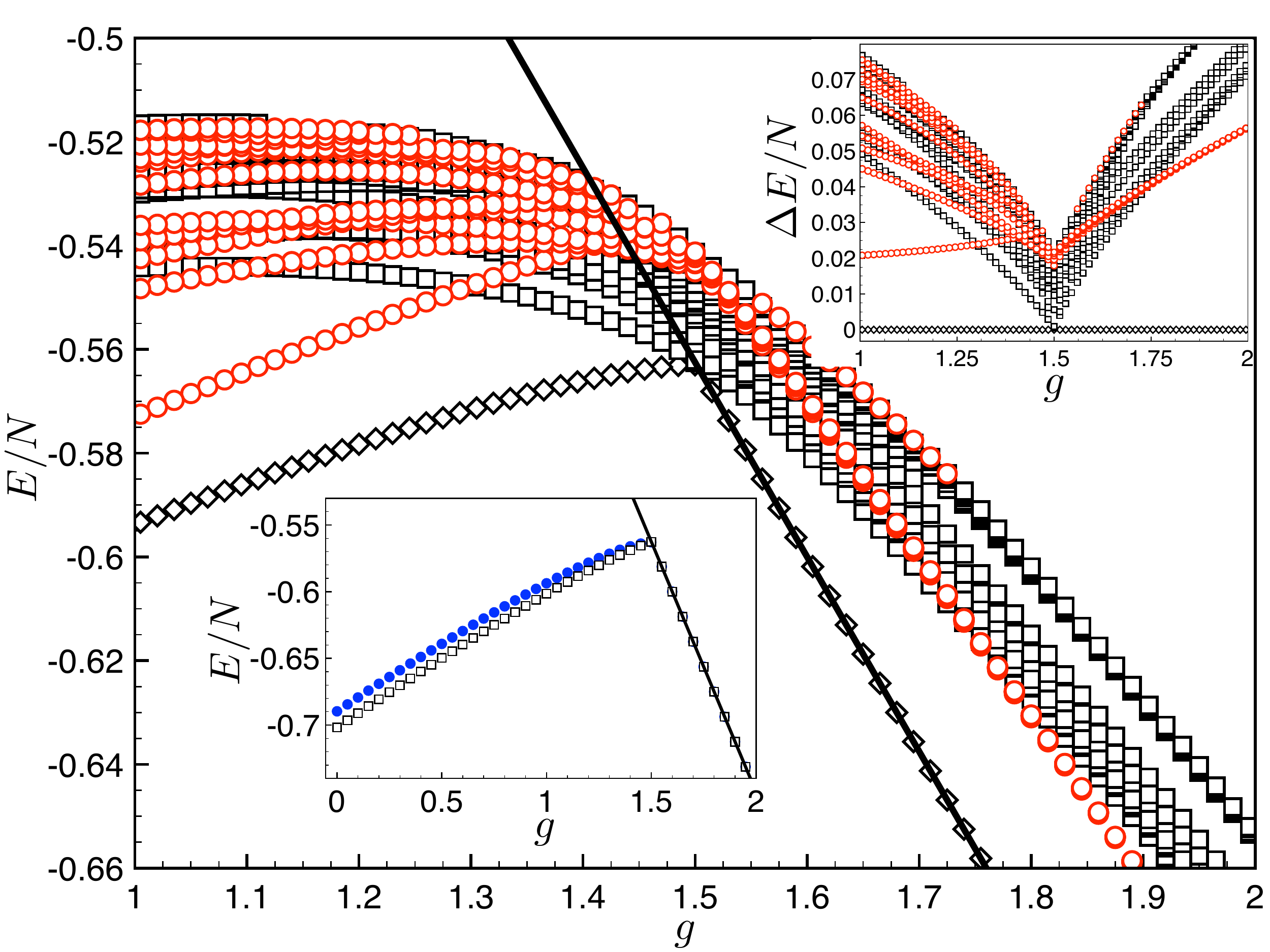}
\caption{(Color online) Per site energy spectrum for lowest  SSM excitations obtained from exact diagonalization of clusters with $N=24$ spins as 
a function of $g=J_2/J_1$. Black diamonds label ground-state energies, black squares label excited states having zero total spin, and red circles label spin 1 triplet excitations. Upper right inset shows excitation energies relative to the ground-state energy using the same labeling convention used for the main plot. Lower left inset shows calculated per site ground-state energies for clusters with $N=16$ spins (black squares), and $N=24$ spins (solid blue circles).  The black solid line is the energy of the exact dimer-singlet valence bond solid state [see Eq.(~\ref{dimerprod}), i.e., $E/N=-(3/8)g]$.}
\label{Fig2}
\end{center}
\end{figure}

\begin{figure}
\begin{center}
\includegraphics[width=2.7cm,angle=0]{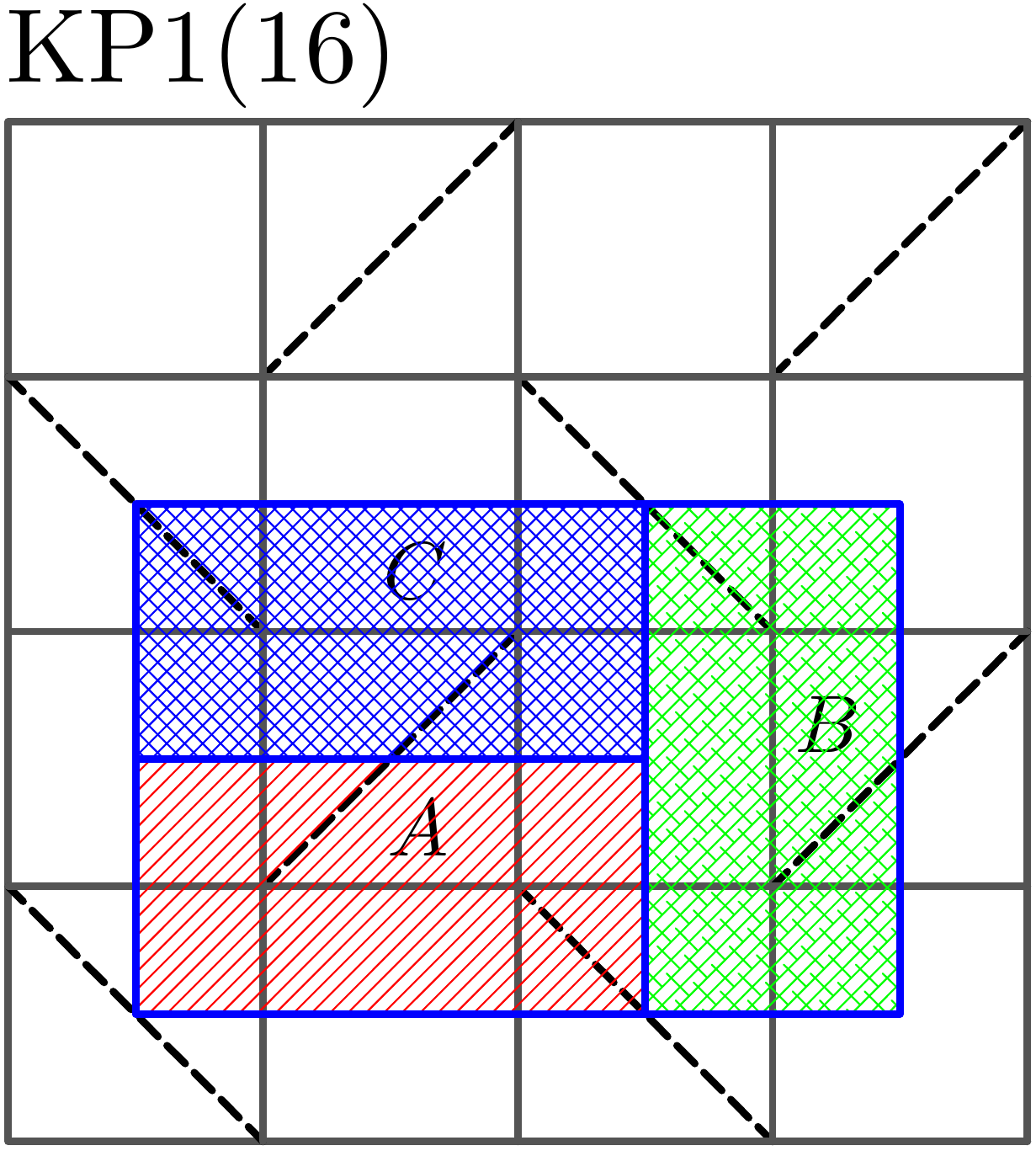}
\includegraphics[width=2.7cm,angle=0]{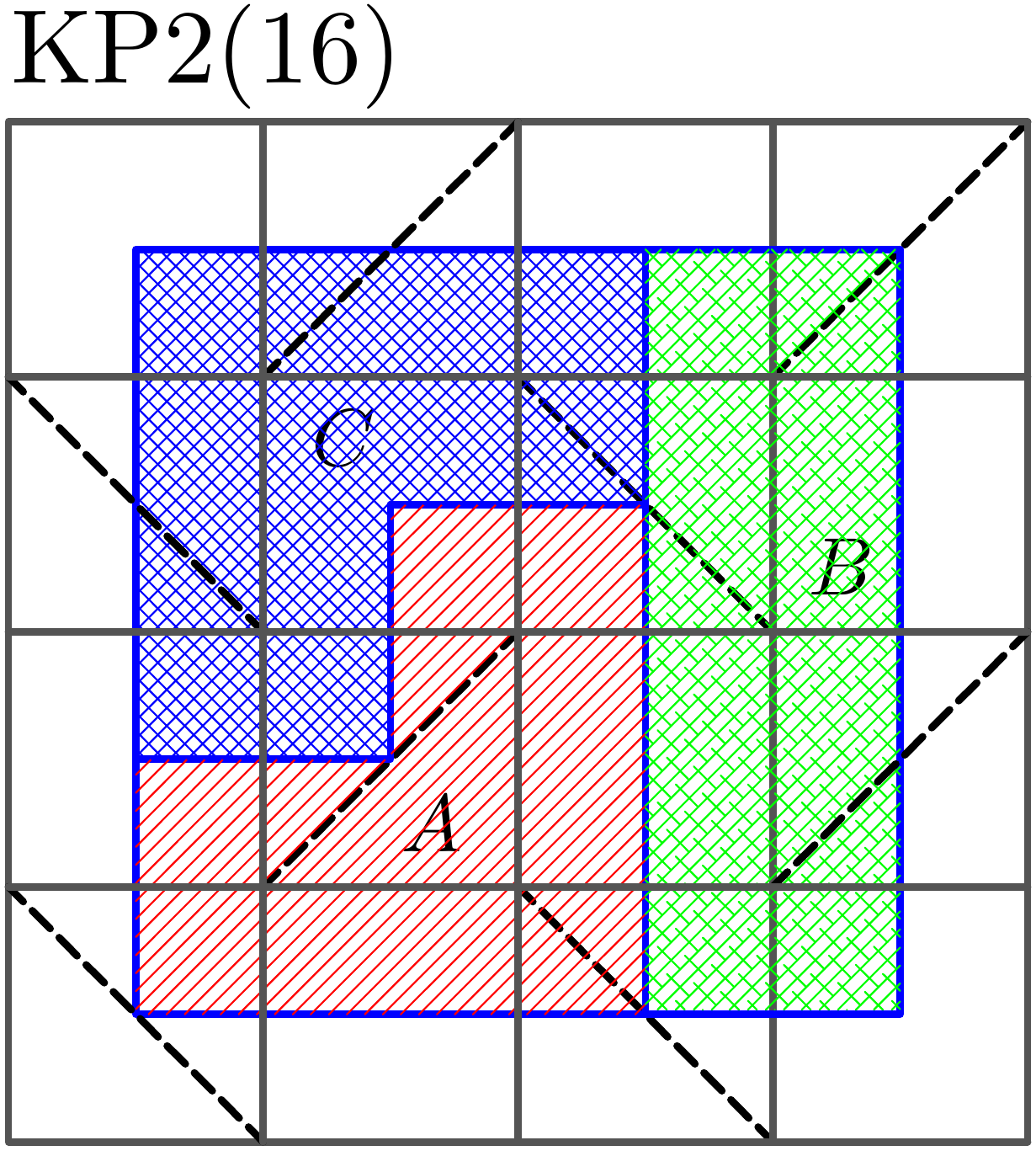}\\
\includegraphics[scale=.1685,angle=0]{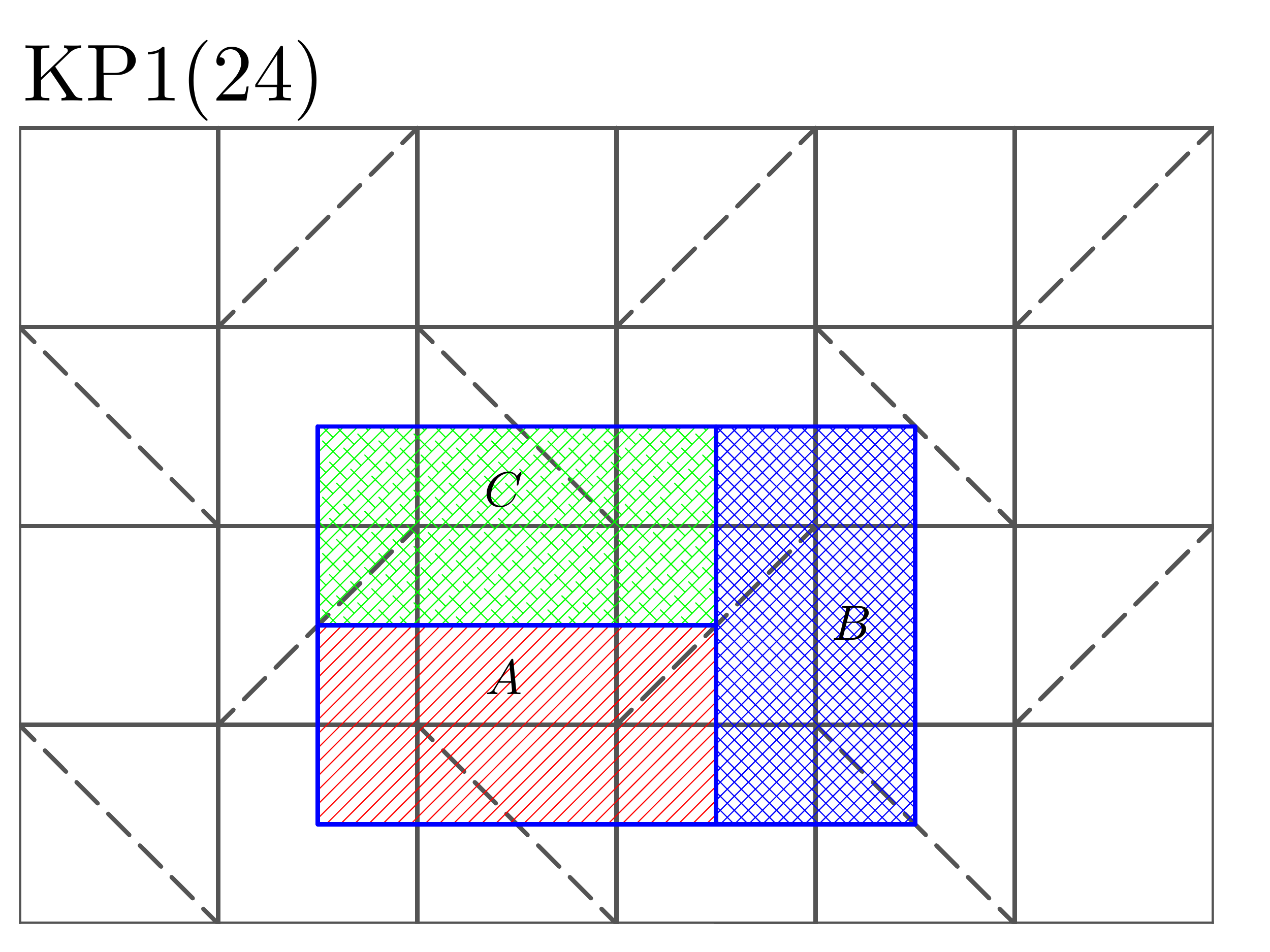}
\includegraphics[scale=.1685,angle=0]{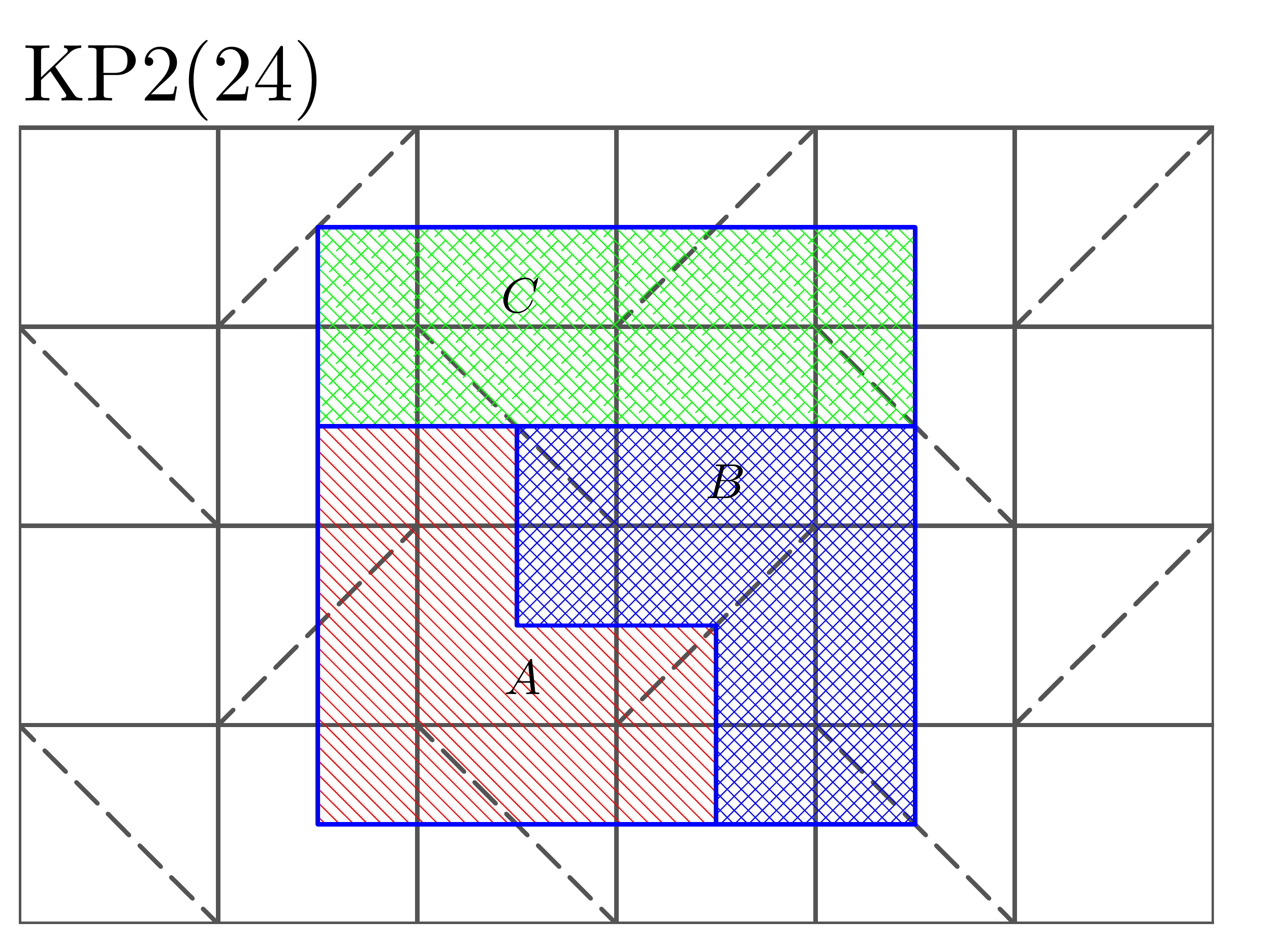}
\includegraphics[scale=.1685,angle=0]{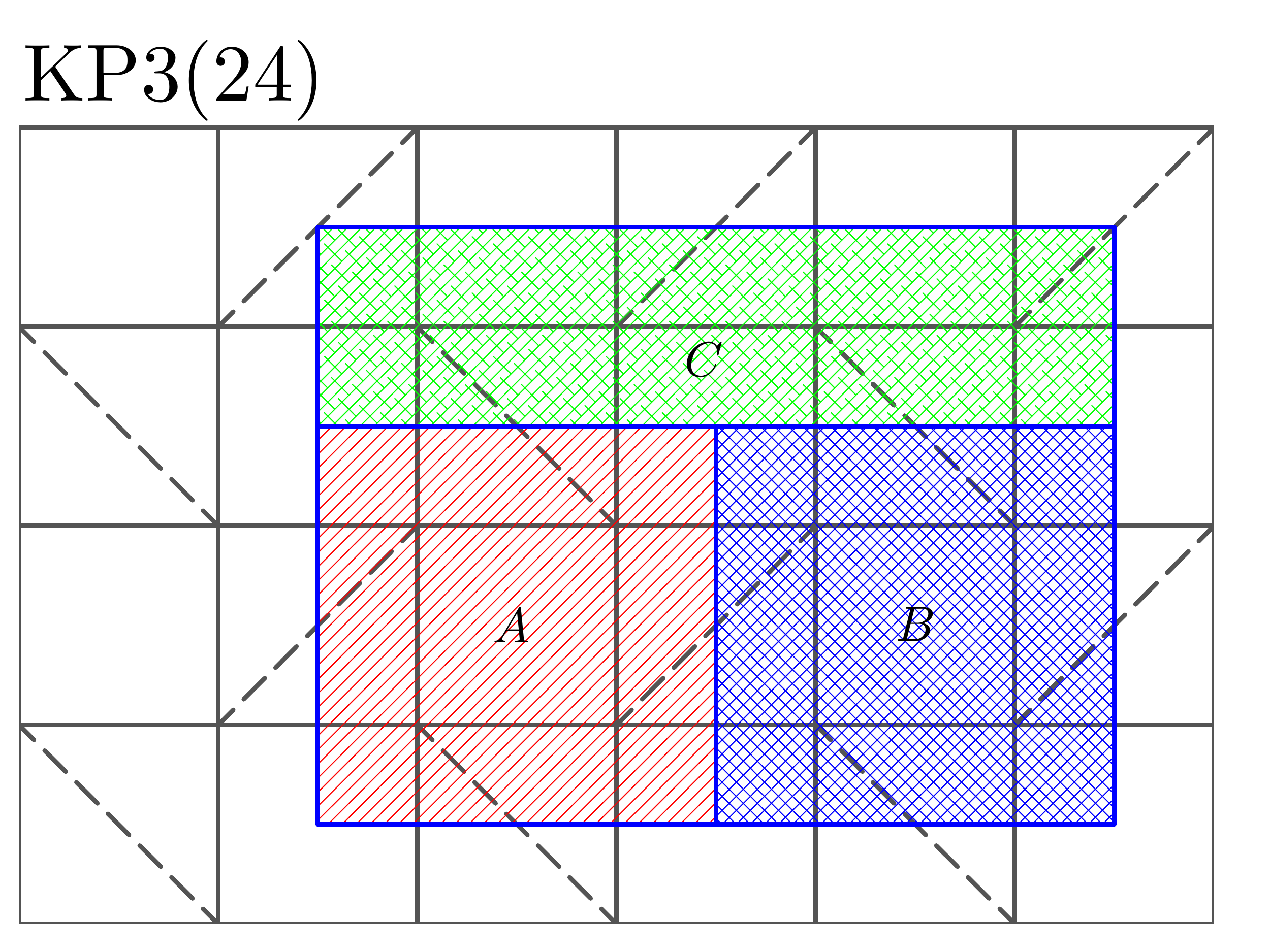}
\includegraphics[scale=.1685,angle=0]{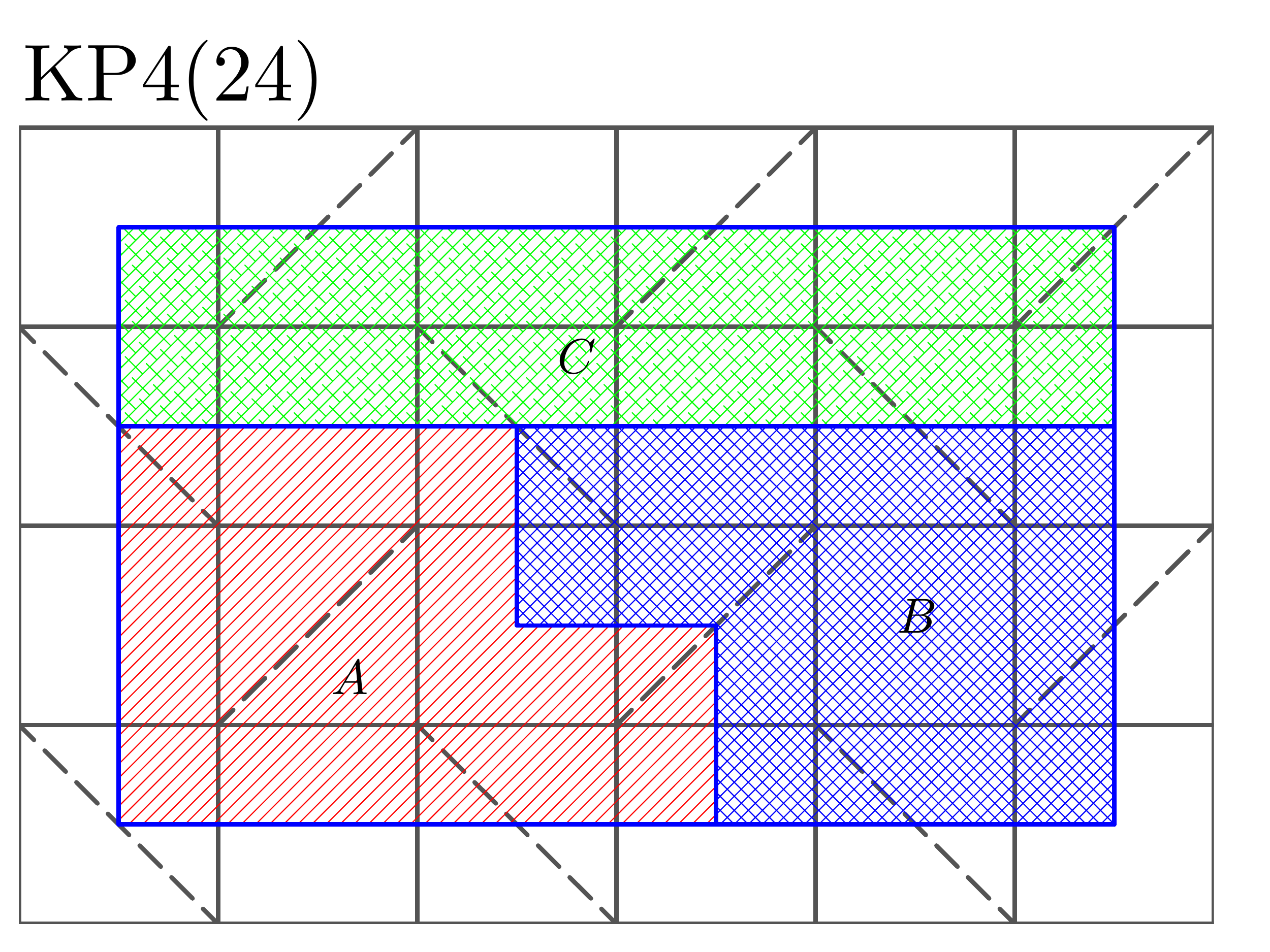}
\caption{(Color online) Kitaev-Preskill constructions for our 16 [top two labeled KP1(16) and KP2(16)] and 24 [bottom 4 labeled KP1(24) through KP4(24)] -site latices.   The 
Shastry-Sutherland lattice is obviously also depicted in the above plots showing a square lattice with $J_{1}$ strength axial bonds (solid lines), and $J_{2}$ strength diagonal bonds (dashed lines) on every other square.}
\label{16kp}
\end{center}
\end{figure}

We mention the futility of employing quantum Monte Carlo methods to the frustrated regimes of 2D quantum antiferromagnets due to the sign problem. Instead we perform Lanczos exact diagonalization to obtain the ground-state wave function $|\psi_{g}\rangle$ of Eq. (\ref{Ham}) across a wide range of the SSM's parameter space for clusters having $N=16$ (4$\times$4) and $N=24$ (4$\times$6) spins on the torus---we note the dimension of the Hilbert space for the 24-site system is well over a million, with the next larger system's (32 sites) space being nearly a half-billion. The lower left inset in Fig. \ref{Fig2} shows our calculated ground-state energy per site for each cluster size. Values obtained for the smaller 16-spin system and the larger 24-spin system are represented by squares and solid circles, respectively. 

The main plot in Fig. \ref{Fig2} shows the per site energy spectrum for some of the SSM's lowest excitations for $1\leq g\leq 2$. The upper right inset of this figure depicts the energy of these excitations relative to the energy of the ground state. These latter two results are from the larger 24-spin system.  Here we observe the gap closing at $g=1.5$ and the proliferation of singlets under the triplet gap, indicative of a quantum spin liquid~\cite{Mila2000}.  
The location of the apparent quantum critical point between the dimer-singlet VBS and the intermediate liquid phase, very near $g=1.5$, is in agreement with various other studies \cite{Weihong1999, Koga2000, Carpentier2001, Zheng2001, Lauchli2002, Darradi2005, Moukouri2008}.  We note results from the 16-spin system are qualitatively similar to the 24-spin system with the main difference being the number of singlets underneath the triplet gap.

We specifically mention the work of L{\"a}uchli \textit{et al.} \cite{Lauchli2002} whose boson operator studies, exact diagonalization, and dimer-dimer four-point correlation function results concerning the SSM identify a finite region surrounding $g=1.45$ as having a twofold ground-state degenerate plaquette VBS order. According to these authors, this phase is intermediate to the SS dimer-singlet VBS phase and to the N\'{e}el-ordered phase, having first- and second-order transitions to each of these, respectively. They identify the transition point between the dimer-singlet VBS and the degenerate plaquette VBS to be at $g=1.5$, which is consistent with the level crossing witnessed in Fig. \ref{Fig2}. More recently, Corboz and Mila, using iPEPS method, have also found evidence of a long-range ordered plaquette phase lying within $g\in[1.307\,19, 1.481\,48]$ \cite{Corboz2013}.

Finally, we mention the work of Chung \textit{et al.} \cite{Chung2001}, which applies a gauge-theoretic analysis to the possible paramagnetic phases of the SSM having Sp(2$N$) symmetry. Though the majority of their results apply only to this more general case, the authors make it clear that some of their results may still apply in the physical $N=1$ limit, where $\text{Sp(2)}\cong\text{SU(2)}$. In particular, the topologically degenerate phase they identified, which we mentioned at the start of this section, could be present within a narrow range of $g$ \cite{Chung2001}. Such a state, which they describe as being a deconfined spin-liquid, is contiguous to the confined SS VBS state. The transition between these states, they show, is described by a $\mathbb{Z}_{2}$ gauge theory.

\textit{Topological entanglement entropy}. The von Neumann entanglement entropy of a bipartite many-body system, $S_{A}=-\text{Tr}\rho_{A}\text{ln}\rho_{A}$ is a measure quantifying the degree of entanglement between a region $A$ and its compliment $B$. Here, $\rho_{A}$ is the reduced density matrix of $A$ obtained by tracing out from $\rho_{A\cup B}$ the degrees of freedom pertaining to $B$: $\rho_{A}=\text{Tr}_{B}|\psi_{g}\rangle\langle\psi_{g}|$, where $|\psi_{g}\rangle$ is the ground-state wave function \cite{Srednicki1993,Vidal2003,Amico2008,Latorre2009, Peschel2009, Fradkin2009, Haque2009, Furukawa2007, Papanikolaou2007, Jiang2012}.

Topological entanglement entropy (TEE) was introduced by Kitaev and Preskill (KP) \cite{Kitaev2006}, and Levin and Wen \cite{Levin2006} as an unambiguous identifier of topological order  within globally entangled ground states. It arises as a negative correction to the area law \cite{Srednicki1993} for entanglement entropy
\begin{equation}\label{arealaw}
S_A=\alpha L-\gamma +\mathcal{O}(1/L),
\end{equation}
where $L$ is the length of the \textit{smooth} boundary between regions, 
and $\alpha$ is a non-universal factor relating to short-range correlations across the boundary. The last term in Eq.~(\ref{arealaw}) represents higher-order terms which vanish in the $L\rightarrow\infty$ limit. The subleading $S_{\text{topo}}\equiv-\gamma<0$ term is the TEE, a distinctive, universal constant characterizing global entanglement in the ground-state wave function \cite{Kitaev2006}. In a topologically ordered state, $S_\text{topo}=-\text{ln}\,\mathcal{D}$, where $\mathcal{D}$ is the total quantum dimension, given by $\mathcal{D}=\sqrt{\sum_{i}d_{i}^{2}}$, with $d_{i}$ being the $i$th quasiparticle's quantum dimension \cite{Kitaev2006, Furukawa2007, Grover2013}.  
 
For nontopologically ordered phases, the area law is faithfully observed with $S_{\text{topo}}=0$. Positive constant contributions of a nontopological nature can also arise in Eq. \ref{arealaw}. These result from Goldstone modes in a non-degenerate ground-state belonging to a symmetry-broken phase \cite{Jiang2012, Helmes2014, Selem2013}. Such contributions (should they arise) will not affect our ability to detect TEE, since they can only arise within nontopologically ordered regimes. Finally, nontopological contributions arising from sharp boundary edges between bipartite regions \textit{may} play the role of ``noise" and interfere with our ability to clearly detect $-\gamma$ within a topologically ordered phase \cite{Helmes2014, Selem2013, Metlitski2011, Kallin2011}. 

Considerate of these facts, we proceed in employing KP plural area constructions, which work to neutralize corner effects by canceling out all boundary contributions to the entanglement entropy. In the large system limit, the possible values $S_{\text{topo}}$ may assume will be fully determined by the physics of the system, and will be entirely independent of the particular region geometries used for calculating entanglement entropy. The identification of a negative value for $S_{\text{topo}}$ can then  be interpreted as positive confirmation for the existence of a topologically ordered state \cite{Jiang2012}. In what follows we describe our approach to identifying such states within a finite region of the SSM's bond-strength parameter space.

\textit{Numerical Results}---According to Kitaev and Preskill \cite{Kitaev2006}, the TEE of a topologically ordered phase can be extracted by dividing the plane of the lattice into four contiguous regions $A$, $B$, $C$, and $D$, three of which are embedded within the fourth (say $D$), and taking the following linear combination of von Neumann entropies
\begin{equation}\label{KP}
S_{\text{topo}}=S_{A}+S_{B}+S_{C}-S_{AB}-S_{BC}-S_{CA}+S_{ABC},
\end{equation}
where $S_{XY}$ and $S_{XYZ}$ are the entanglement entropies of the combined regions $X\cup Y$ and $X\cup Y\cup Z$, respectively.

We choose our KP areas to be polygons whose corners form either a 90$\degree$ or 270$\degree$ angle. We choose unit vectors $\hat{a}_{x}=\hat{a}_{y}$, which correspond to the $g=0$ length of a primitive square's edge on the underlying Heisenberg lattice. Letting $a_{x}=a_{y}=1$, the perimeter of a combined $ABC$ area is $L=2L_{x}+2L_{y}$, where $L_{x}$ and $L_{y}$ are the number of sites along respective $x$ and $y$ directions within a given area. 

To ensure that we achieve sufficiently small $\xi(g)/L_s$ \cite{Papanikolaou2007, Grover2013} [where $\xi(g)$ is the $g$ dependent correlation length and $L_s$ is the perimeter of a KP subregion] we increase $L$  by using various composite KP area sizes, and tune the parameter $g$ to minimize the correlation length $\xi(g)$ along the long-range ordered to highly frustrated paramagnetic phase transition.

We wish to see how $S_{\text{topo}}$ varies with respect to $g$ and the size of the composite KP area $ABC$. To this end, we form on the 16-site lattice two sets of $A$, $B$, and $C$ subregions labeled KP1(16) and KP2(16), as shown in Fig. \ref{16kp}. For each of these constructions the perimeter of each subregion crosses a definite, and relatively varied, number of axial and diagonal bonds. Each such crossing occurs at the midpoint of any given bond. We similarly construct KP regions for the 24-site SS lattice, also shown in Fig. \ref{16kp}.

\begin{figure}
\begin{center}
\hspace*{-.6cm}
\includegraphics[width=8.5cm,angle=0]{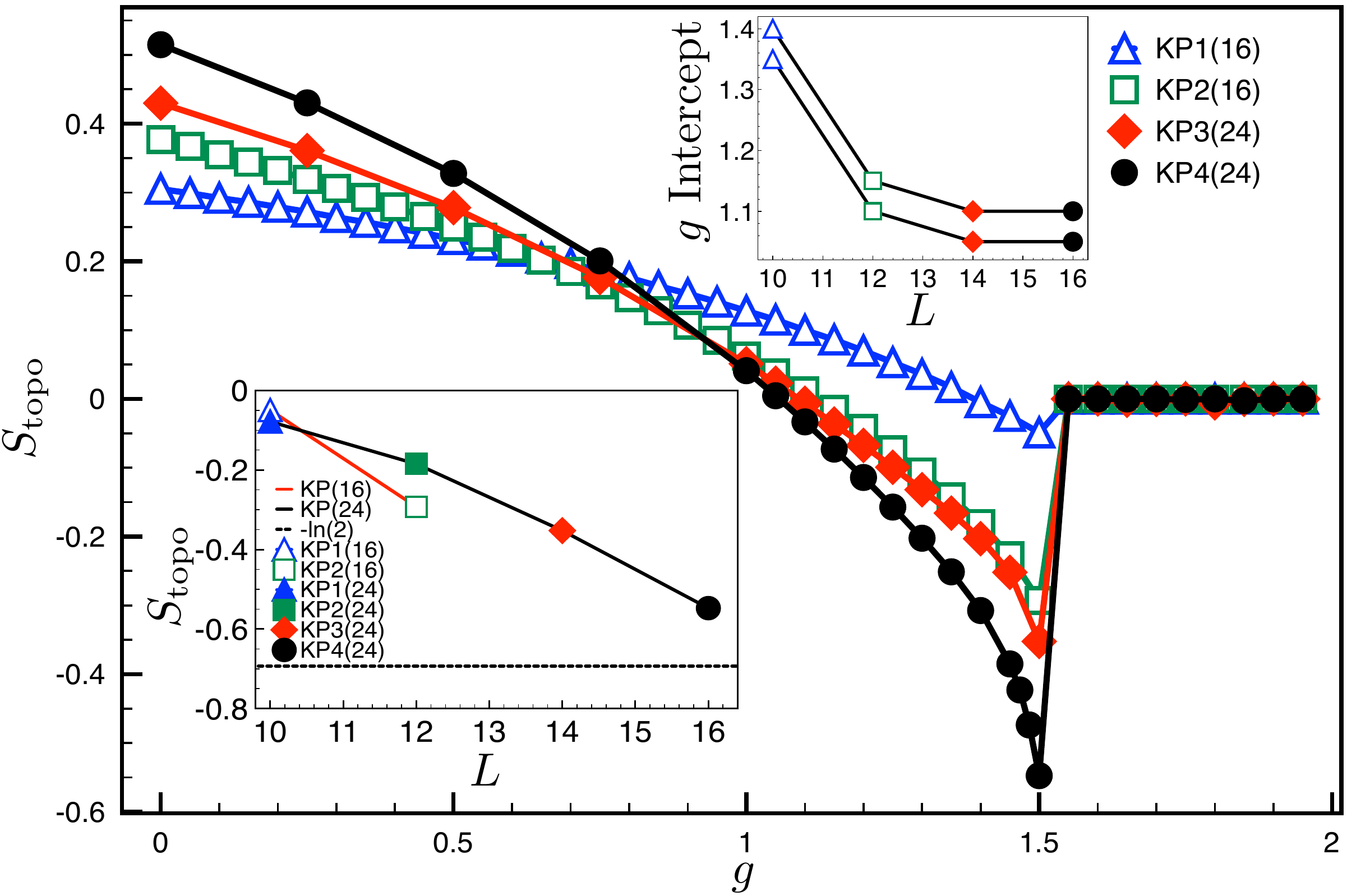}
\caption{(Color online) Topological entanglement entropy for  various Kitaev-Preskill constructions, $S_\mathrm{topo}$, as a function of $g=J_{2}/J_{1}$. See  Eq.~(\ref{KP}) and Fig.~\ref{16kp} for details.  The lower left inset shows $S_{\text{topo}}$ at $g=1.5$ for various Kitaev-Preskill constructions as a function of the bipartite boundary length $L$ (the perimeter of the combined $ABC$ area). The upper right inset shows the neighborhood within which the $g$ intercept of the main plot lies (region between each curve), as a function of $L$.}
\label{fig-tee}
\end{center}
\end{figure}
Figure \ref{fig-tee} shows the evolution of $S_{\text{topo}}$ across varying $g\in[0,2)$, for various system and area sizes. Immediately, we notice that $S_{\text{topo}}=0$ exactly for $g>1.5$ as expected for any short-range entangled state, such as the one described by the product in Eq. (\ref{dimerprod}). Below this regime we expect the combination in Eq.~(\ref{KP}) to yield either a positive nontopological correction within a symmetry-broken phase, or a 
negative topological value within a topologically ordered one.  

Each of the four plots in Fig. \ref{fig-tee} demonstrate clear regimes (below $g<1.5$) within which $S_{\text{topo}}$ is either positive or  negative. The location of the transition point between positive and negative $S_{\text{topo}}$ values (the $g$ intercept; see upper right inset in Fig. \ref{fig-tee}) for a given curve shifts leftwardly from around $g=1.35-1.40$ to $g=1.05-1.10$, as the perimeter of the combined $ABC$ area $L$ increases. The transition points appear to converge near these latter values for ever larger $L$.

In an infinite system one would expect a positive $S_{\text{topo}}$ to suddenly drop to $-\gamma$ along the symmetry-broken to spin-liquid phase transition.  Our finite sized systems and area sizes prevent us from seeing such suddenness; however, Fig. \ref{fig-tee}  clearly shows a strong tendency in this direction. For fixed $g$ within the $S_{\text{topo}}<0$ regime, $S_{\text{topo}}$ decreases as a function of $L$, while it increases along with greater $L$ within the $S_{\text{topo}}>0$ regime.

As $g\rightarrow 1.5$ from the left, $S_{\text{topo}}$ decreases monotonically for each of the curves, converging toward respective minima. Each curve then suddenly jumps to zero upon passing into the $g>1.5$ regime. The lower left inset in Fig. \ref{fig-tee} shows $S_{\text{topo}}$ as a function of $L$ at this transition point, which corresponds to the widely suspected quantum critical value $g=1.5$. Here, $S_{\text{topo}}$ monotonically decreases with $L$ consistent with an approach to the $-\text{ln}(2)$ $\mathbb{Z}_2$ topological value (horizontal line) for larger area sizes on the 24-spin cluster--however, we note that this behavior is merely consistent with the value for a $\mathbb{Z}_2$ quantum spin liquid and could, in fact, be approaching a different value corresponding to a different topological order.  Though we expect similar exact diagonalization calculations on larger systems would yield better convergence, such an attempt is impractical given the extremely large size of the Hilbert space for the next larger 32-spin system, as mentioned previously.  While it is possible to find the ground state for the 32-spin system (see L{\"a}uchli \textit{et al.}~\cite{Lauchli2002}) the calculation of the reduced density matrices in order to find the TEE using the Kitaev-Preskill construction would be prohibitive.

\textit{Conclusion}---We have shown via Kitaev-Preskill constructions that the ground-state wave function for the 2D Shastry-Sutherland model is topologically ordered at $g=1.5$, with this state extending below this point for some finite neighborhood, possibly extending down to near $g=1.1$. This was done by obtaining $S_{\text{topo}}<0$ within much of the latter domain, specifically for increasing $L$, and suggesting convergence with expected values for $S_\mathrm{topo}$, at $g=1.5$, consistent with various topologically ordered states (for example, see Ref.~[\onlinecite{Grover2013}]). Importantly, our result makes it highly unlikely that the intermediate state is a symmetry-broken product state, having some finite local order parameter, such as a plaquette state \cite{Corboz2013, Lou2012}.  Such a state would yield a non-negative $S_{\text{topo}}$ result. Additionally, any candidate topologically ordered state with $S_{\text{topo}}$ greater than our lowest attained value ($S_{\text{topo}}=-0.547\,510\,6$) is unlikely to describe the intermediate state.

The discrete jump at $g=1.5$ in Fig. \ref{fig-tee} clearly identifies the existence of a first-order transition between the topologically ordered state and the SS dimer-singlet VBS; the latter state exhibiting the expected $S_{\text{topo}}=0$ result for short-range entangled states. $S_{\text{topo}}$'s continuous change of sign around $g=1.1$, for larger $L$, leads us to suspect the transition here to be of second order, in agreement with most prior studies \cite{Koga2000, Carpentier2001, Chung2001, Zheng2001, Lauchli2002, Miyahara2003}. 

The description of the SSM's phases here presented uses topological entanglement entropy to support prior suspicions and results (see Fig. \ref{Fig1}). While we bring clarity to the long-standing issue regarding this model's intermediate phase, confirming its exotic nature, the exact global entanglement patterns exhibited by the SSM within the topologically ordered regime are still not entirely known; further work in this direction is still needed. Applying approximation methods to finding TEE on larger SSM clusters may help to shed further light. 

\textit{Acknowledgments}---We acknowledge California State University Long Beach Office of Research and Sponsored Programs and D. C. R. thanks the 
Office of Research and Sponsored Programs Summer Research Assistantship.  We benefitted from conversations and comments from Andreas Bill, Thierry Jolicoeur, Vito Scarola, and Sriram Shastry.  We thank Roderich Moessner for helpful comments improving the manuscript.

\end{document}